\documentstyle[12pt]{article}
\newcommand{\eq}{\begin{equation}}
\newcommand{\en}{\end{equation}}
\newcommand{\eqn}{\begin{eqnarray}}
\newcommand{\enn}{\end{eqnarray}}
\newcommand{\CR}{\nonumber \\}
\newcommand{\del}{\partial}
\newcommand{\lax}{{\cal L}}
\newcommand{\orl}{{\cal M}}

\begin{document}

\renewcommand{\thefootnote}{\fnsymbol{footnote}}
\begin{titlepage}
\null
\begin{flushright}
hep-th/9502029
\end{flushright}
\vspace{2.0cm}
\begin{center}
{\Large \bf
Topological Strings with Scaling Violation \\
and Toda Lattice Hierarchy
\par}
\lineskip .75em
\vskip 3em
\normalsize
{\large Hiroaki Kanno}
\vskip 1.5em
{\it Department of Mathematics, Faculty of Science, Hiroshima University, \\
Higashi-Hiroshima 724, Japan}
\vskip 1.5em
 and
\vskip 3em
\normalsize
{\large Y\H uji Ohta}
\vskip 1.5em
{\it Department of Physics, Faculty of Science, Hiroshima University, \\
Higashi-Hiroshima 724, Japan}
\end{center}
\vskip3em

\begin{abstract}
We show that there is a series of topological string theories whose
integrable structure is described by the Toda lattice hierarchy.
The monodromy group of the Frobenius manifold for the matter
sector is  an extension of the affine Weyl group $\widetilde W
(A_N^{(1)})$ introduced by Dubrovin. These models are generalizations
of the topological $CP^1$ string theory with scaling violation.
The logarithmic Hamiltonians generate flows for the puncture
operator and its descendants. We derive the string equation
from the constraints on the Lax and the Orlov operators.
The constraints are of different type from those for the
$c=1$ string theory. Higher genus expansion is obtained
by considering the Lax operator in matrix form.
\end{abstract}
\end{titlepage}
\renewcommand{\thefootnote}{\arabic{footnote}}
\setcounter{footnote}{0}
\baselineskip=0.7cm

\section{Introduction}

\renewcommand{\theequation}{1.\arabic{equation}}\setcounter{equation}{0}

Integrable structure of topological string theories has
attracted much attention in recent years \cite{Dij} \cite{Dub}.
Topological string theory is a kind of two dimensional cohomological
field theory with topological matter coupled to topological gravity.
As a cohomological field theory, topological string theory gives
topological correlation functions among BRST cohomology classes.
A generating function of these correlation functions is
the free energy of the theory (the logarithm of the partition function).
The integrability of topological string theory means that the partition
function
is a tau function of some integrable hierarchy.
Furthermore, the partition function is uniquely characterized in terms of
the so-called string equation.
One of the motivations for investigating such a integrability of topological
string theory
is that the non-critical string theory has several common
algebraic structures with topological string theory.
Indeed, any string theory has more or less characteristic
properties as topological field theory.

An important class of topological matter theory is
topological conformal matter which can be obtained by
twisting any $N=2$ super conformal model.
Taking the minimal $N=2$ superconformal model of ADE type,
which allows the Landau-Ginzburg description,
one can construct topological string theory whose integrable
structure is identified as the generalized KdV hierarchy,
or the Drinfeld-Sokolov hierarchy of ADE type \cite{DVV}.
Topological matter theory usualy has a finite number of
physical operators, called (chiral) primary fields.
After coupling to topological gravity, an infinite series
of gravitational descendants appears for each primary
field. These descendants correspond to
infinite number of commuting Hamiltonians of
integrable hierarchy.

Recently it has been shown that there exist
physically more interesting topological strings which are integrable.
The $c=1$ string theory is an important example \cite{Kle} \cite{GM}.
It is originally defined as a non-critical string theory and
has physical scattering process among tachyons.
Therefore, it is an astonishing fact that the $c=1$ string
theory is topological \cite{MV} \cite{AGSY} \cite{IsK} \cite{GhM}
and integrable \cite{DMP} \cite{HOP}.
Another example is the topological $CP^1$ model \cite{Wit} \cite{DW}
\cite{EYCP}.
This is also quite intersting, since it
opens a possibility of investigating quantum geometry
of the target space, such as quantum cohomology ring,
from the view point of integrable systems.
The Toda lattice hierarchy is the integrable structure
behind both examples. In this paper we will show that
there is a series of topological string theories whose integrable
structure is described by the Toda lattice hierarchy.
The difference of the models comes from the
constraints imposed on the Lax operators, or in other words, the string
equation.
In \cite{Ta1}, \cite{EK}, it is shown that the string equation of the $c=1$
string
theory is given by the constraints;
\eq
\lax~=~\overline\orl~\overline\lax~, \quad
\overline\lax^{-1}~=~\orl \lax^{-1}~,  \label{constA}
\en
on the Lax operators $\lax, \overline\lax$ and the Orlov operators
$\orl, \overline\orl$ of the (dispersionless) Toda lattice
hierarchy. (For the definition of these objects, see section 4.)
It has been established that the constraints (\ref{constA}) imply
the $W_{1+\infty}$ constraints of the $c=1$ theory which
uniquely determine all the tachyon correlation functions
at the self-dual radius \cite{DMP},\cite{HOP}, \cite{GIM}. The string equation
in full genus is discussed in \cite{Nak}. Quite recently the string equation
of the deformed $c=1$ string in the black hole background is also formulated
in the framework of the Toda lattice hierarchy \cite{NTT}.

Topological string theories we will consider in this paper have
a different type of the constriants;
\eq
{\lax^N \over N}~=~\overline\lax^{-1}~, \quad
\orl \lax^{-N}~=~-\overline\orl~\overline\lax~, \quad
(N= 1,2. \cdots).  \label{constB}
\en
We note that similar constraints have been considered by Takasaki
in connection with the topological strings of $D$ type \cite{Ta2}.
Contrary to the constraint of the first type which allows infinitely many
primary fields, the second type is quite restrictive on the number of
the primaries which survive. After imposing the constraints (\ref{constB}), we
have only
$(N+1)$ primary fields. The case $N=1$ corresponds to the topological
$CP^1$ model. Hence we have a series of topological string theories $(N\geq 2)$
which generalize it. One of remarkable features in this type of topological
string theory is that the flows for the puncture operator and its descendants
are generated by the logarithmic Hamiltonians, which was first observed
in \cite{EYCP}. The standard Hamiltonians in the Toda lattice hierarchy are of
polynomial type. The logarithmic Hamiltonians involve the logarithm
of the Lax operator. Such logarithmic Hamiltonians are well-defined
only after the constrains of type (\ref{constB}) are imposed. Thus, they are
peculiar to our models with the constraints (\ref{constB}).

In \cite{BX1}, it is shown that the constraints (\ref{constA}) arise quite
naturally
from the two matrix model without continuum limit.
We easily see that the constraints (\ref{constB}) for $N=1$ reduces the two
matrix model to the one matrix model \cite{EYCP}. Though we believe that
there is an appropriate matrix model realization for $N\geq 2$,
we have not found it at present.

In the following we will be mainly concerned with an example
with three primaries for simplicity. The primary fields are
the puncture operator $P$, one marginal operator $R$
and one relevent operator $Q$. (In general models,
we will have several relevant operators but only
one marginal operator.)
Our model looks like a kind of \lq\lq fusion" of
the topological minimal model and the topological
$CP^1$ model. The marginal operator comes from
the $CP^1$ sector, while the relevant operator
is similar to that of topological $A_N$ model.
Like the topological $CP^1$ model, matter sector
of our model does not have conformal invariance,
or its \lq\lq effective" first Chern class is non-vanishing.
In fact the topological correlation functions have
terms which may be interpreted as the instanton corrections,
though their geometrical meaning is somewhat
obscure.

The present paper is organized as follows.
In the next section we take a topological matter theory
first introduced by Dubrovin from his viewpoint of
the Frobenius manifold and
compute topological correlation functions using the topological
recursion relation at tree level. The Lax formalism
which implies all the result of tree level correlation functions
is introduced in section 3.  As we have seen in the topological
$CP^1$ model, the flows corresponding to the descendants
of the puncture operator are generated by the Hamiltonians
including the logarithm of the Lax operator.  In section 4,
the integrable structure behind this Lax formalism
is shown to be the Toda lattice
hierarchy with a special constraint on the Lax operators.
We will derive the string equation of the theory from the constraints
on the Lax and the Orlov operators. This approach is similar
to the one we have used for the $c=1$ string theory.
Section 5 is devoted to the higher genus structure.
We will examine the genus expansion by using
the Lax operator in matrix form.

\newpage
\section{Tree Level Correlation Functions}

\renewcommand{\theequation}{2.\arabic{equation}}\setcounter{equation}{0}

The topological string theory we will consider throughout
the present paper is constructed from a topological matter
introduced by Dubrovin.
(See Exercise 4.3 and Example 5.5 in \cite{Dub})
Before coupling to gravity the tree level free energy takes the form;
\eq
F(t)~=~{1\over 2}(t_1)^2 t_3 + {1\over 2}t_1(t_2)^2
-{1\over 24}(t_2)^4 + t_2 e^{t_3}~. \label{free}
\en
This model was discovered by considering the monodromy of
the Frobenius manifold associated with topological matter.
The monodromy of (\ref{free}) is an extention of the affine Weyl group
$\widetilde{W}(A_2^{(1)})$.  We can also find this model in the classification
program of $N=2$ theories by Ceccoti and Vafa \cite{CV}. It is among three
primary
models\footnote{We thank S.-K.~Yang for pointing it out.}.
The topological matter with the
monodromy $\widetilde{W}(A_1^{(1)})$ gives the same topological string theory
as the topological $CP^1$ model. There are a series of topological
matter theories with the monodromy $\widetilde{W}(A_N^{(1)})$, which
generalizes the topological $CP^1$ model. (Another more geometrical
generalization is, of course, the topological $CP^n$ model, or the
topological Grassmannian model. But their integrable structure is
still an open problem.)
The free energy has an important property of generalized
homogeneity;
\eq
(3 - \lax_E) F(t)~=~F(t)~,
\en
where the Euler vector field $E$ is given by
\eq
E~=~t_1 {\del\over \del t_1} + {1\over 2}t_2 {\del\over \del t_2}
 + {3\over 2} {\del\over \del t_3}~.
\en
We see that the dimension of this model is one.
The parameter $t_3$ is a marginal one. The coefficient $3/2$
of $\del/ \del t_3$ can be regarded as the \lq\lq effective"
first Chern class in counting the ghost number anomaly.

The primary couplings $t_i (i=1,2,3)$ are chosen to be
flat coordinates.  Let us introduce the primary
operators $P,Q,R$ conjugate to these couplings. Accordingly we will
change the notation as $t_{0,P}\equiv t_1, t_{0,Q}\equiv t_2,
t_{0,R}\equiv t_3$.
Then the basic two point functions, or the order parameters are
\eq
\langle PP \rangle \equiv u(t) = t_{0,R}, \quad
\langle PQ \rangle \equiv v(t) = t_{0,Q}, \quad
\langle PR \rangle \equiv w(t) = t_{0,P}.
\en
On the small phase space the order parameters take
the simple form; $u(t)=t_{0,R}, v(t)=t_{0,Q}, w(t)=t_{0,P}$.
After coupling to topological gravity, they do not have
such simple forms any more. However, the topological recursion
relation to be introduced shortly guarantees that any topological
correlation functions are expressed in terms of these order
parameters. That is, the dependence on the descendant couplings
is only through the dependence of $u(t),v(t),w(t)$.
In \cite{DW}, such relations for two point functions;
\eq
\langle AB \rangle (t)~=~R_{AB}(u(t),v(t),w(t))~
\en
are treated as the constitutive relations of the theory.
Other primary two point functions, which are the first examples
of the constitutive relations,  are
\eq
\langle QQ \rangle = w -{1\over 2}v^2, \quad
\langle QR \rangle = e^u, \quad
\langle RR \rangle = v e^u.
\en

Topological correlation functions  imply the following
primary flow equations;
\eqn
{\del u\over \del t_{0,Q}}&=&v'~, \quad
{\del v\over \del t_{0,Q}}=\bigl( w-{1\over 2}
v^2\bigr)'~, \quad
{\del w\over \del t_{0,Q}}=(e^u)'~, \CR
{\del u\over \del t_{0,R}}&=&w'~, \quad
{\del v\over \del t_{0,R}}=(e^u)'~, \quad
{\del w\over \del t_{0,R}}~=~\big(ve^u\bigr)'~.
\enn
We will identify the primary field $P$ as the puncture operator,
which implies an identification of $t_{0,P}$ as the space variable $x$.
The prime denotes the derivative with respect to $t_{0,P}\equiv x$.
As a consequence of the primary flows we recognize the dispersionless
limit of the 2D Toda lattice equation;
\eq
{\del^2 u \over \del t_{0,Q} \del t_{0,R}}~=~
{\del^2\over \del x^2}e^u~.
\en

After coupling to topological gravity, there appear
hierarchies of the gravitational
descendants in the non-trivial BRST cohomologies. For each
primary $\Phi_\alpha$, its $n$-th descendant is denoted by
$\sigma_n (\Phi_\alpha),~(n=0,1,\cdots)$.
Using the topological recursion relation at genus zero,
\eq
\langle \sigma_n(\Phi_\alpha) XY\rangle~=~
n\langle \sigma_{n-1}(\Phi_\alpha) \Phi_\beta
\rangle\eta^{\beta\gamma}\langle \Phi_\gamma XY\rangle~,
\en
we can compute the topological
corration functions including the descendants.
 The metric is $\eta^{PR}=\eta^{QQ}=1$.
As was shown by Dijkgraaf and Witten \cite{DW}, the topological
recursion relation at genus zero is deeply
connected to the integrable hierarchy in the dispersionless limit.
We can also see this crucial relation from the viewpoint of
the Gauss-Manin connection \cite{Dub} \cite{Kri} \cite{BV} \cite{EYYP}.

Topological correlation functions for the first descendants are;
\eq
\langle \sigma_1(P) P \rangle~=~
uw + {1\over 2}v^2~,\quad
\langle \sigma_1(Q) P \rangle~=~
vw - {1\over 6}v^3 + e^u~,\quad
\langle \sigma_1(R) P \rangle~=~
{1\over 2}w^2 + ve^u~.
\en
\eq
\langle \sigma_1(P) Q \rangle~=~
(u-1)e^u + vw - {1\over 3}v^3~, \quad
\langle \sigma_1(Q) Q \rangle~=~
ve^u + {1\over 2}w^2 - {1\over 2}wv^2
+{1\over 8}v^4~, \quad
\langle \sigma_1(R) Q \rangle~=~
we^u~.
\en
\eq
\langle \sigma_1(P) R \rangle~=~
uve^u + {1\over 2}w^2~, \quad
\langle \sigma_1(Q) R \rangle~=~
(w + {1\over 2}v^2)e^u~, \quad
\langle \sigma_1(R) R \rangle~=~
wve^u + {1\over 2}e^{2u}~.
\en
To obtain the above result we have integrated once with respect to
$t_{0,P}$ assuming the absence of the integration constants.
It is clear that we can continue this procedure of obtaining
the constitutive relations recursively on the degree of the
descendants. The results on the second descendants, for example, are;
\eqn
\langle \sigma_2(P) P \rangle&=&
uw^2 + wv^2 -{1\over 6}v^4
+ 2v(u-1)e^u~,\CR
\langle \sigma_2(P) Q \rangle&=&
\bigl( 2w(u-1) + v^2\bigr)e^u + vw^2
+{2\over 15}v^5 -{2\over 3}v^3w~, \CR
\langle \sigma_2(P) R \rangle&=&
{1\over 3}w^3 + \bigl( 2uvw + {1\over 3}v^3\bigr)e^u
+ \bigl( u-{3\over2}\bigr)e^{2u}~.
\enn
\eqn
\langle \sigma_2(Q) P \rangle&=&
vw^2 - {1\over 3}v^3 w  + {1\over 20} v^5 +(2w+v^2)e^u~,\CR
\langle \sigma_2(Q) Q \rangle&=&
{1\over 3}w^3 - {1\over 2}v^2w^2 + {1\over 4}v^4 w - {1\over 24}v^6
+(2vw -{1\over3}v^3)e^u + e^{2u}~,\CR
\langle \sigma_2(Q) R \rangle&=&
(w^2 + v^2 w - {1\over 12}v^4)e^u + 2v e^{2u}~.
\enn
\eqn
\langle \sigma_2(R) P \rangle&=&
{1\over 3}w^3 + 2vw e^u + {1\over 2}e^{2u}~,\CR
\langle \sigma_2(R) Q \rangle&=&
w^2 e^u + v e^{2u}~,\CR
\langle \sigma_2(R) R \rangle&=&
w^2 v e^u + (w + v^2) e^{2u}~.
\enn

\section{Lax Formalism}

\renewcommand{\theequation}{3.\arabic{equation}}\setcounter{equation}{0}

Topological correlation functions are translated into the
flow equations for $u(t), v(t)$ and $w(t)$.
Let $t_{n,P},t_{n,Q},t_{n,R}$ be the coupling parameters to
$\sigma_n(P),\sigma_n(Q),\sigma_n(R)$, respectively.
Then the $t_{1,P}$ flow equations are
\eq
{\del u\over \del t_{1,P}}~=~\bigl(
uw + {1\over 2}v^2\bigr)'~,\quad
{\del v\over \del t_{1,P}}~=~\bigl(
(u-1)e^u + vw - {1\over 3}v^3\bigr)'~, \quad
{\del w\over \del t_{1,P}}~=~\bigl(
uve^u + {1\over 2}w^2\bigr)'~. \label{pflow0}
\en
Similarly the $t_{2,P}$ flows read
\eqn
{\del u\over \del t_{2,P}}~&=&~\bigl(
uw^2 + wv^2 -{1\over 6}v^4
+ 2v(u-1)e^u\bigr)'~,\CR
{\del v\over \del t_{2,P}}~&=&~\biggl[
\bigl( 2w(u-1)e^u + v^2\bigr)e^u + vw^2
+{2\over 15}v^5 -{2\over 3}v^3w\biggr]'~, \CR
{\del w\over \del t_{2,P}}~&=&~\biggl[
{1\over 3}w^3 + \bigl( 2uvw + {1\over 3}v^3\bigr)e^u
+ \bigl( u-{3\over2}\bigr)e^{2u}\biggr]'~. \label{pflow1}
\enn
We can write down the $t_{n,Q}$ and the $t_{n,R}$ flow
equations in the same manner.

Let us establish the Lax formalism for the above
flow equations. Our choice of the Lax operator is,
\eq
L~=~{1\over 2}p^2 + vp + w + e^u p^{-1}~.
\en
This Lax operator is suggested by the Landau-Ginzburg potential
proposed by Dubrovin. We employ the following Poisson bracket;
\eq
\biggl\{ A(p,x), B(p,x) \biggr\}~=~
p\biggl( {\del A \over \del p}{\del B \over \del x} -
{\del B \over \del p}{\del A \over \del x}\biggr)~.
\en
Given a Hamiltonian or a flow generator $H(p,x)$, the flow equation
for $u(x,t),v(x,t)$ and $w(x,t)$ is defined by
\eq
{\del L \over \del t}~=~\bigl\{ H, L \bigr\}~.
\en
It is natural to try $(L^n)_+$ first as flow generators.
$(\cdot)_+$ means the non-negative power part.
One can easily see that
\eq
n {\del L\over \del t_{n-1,R}}~=~\bigl\{ (L^n)_+, L\bigr\}, \quad
(n=1,2,\cdots)~,
\en
coincides with the flow equations derived from the topological correlation
functions.
{}From the degree counting we expect that
the $t_{0,Q}$ flow is generated by the square root of the Lax operator;
\eq
L^{1/2}~=~{1\over\sqrt2} \biggl[ p +  v + (w-{v^2\over 2})p^{-1}
+ \bigl( e^u -vw +{1\over 2}v^3 \bigr)
p^{-2} + \cdots \biggr]~.
\en
Indeed the odd powers of it correspond to the $t_{n-1,Q}$ flows.
We have checked explicitly;
\eqn
 {1\over\sqrt2}{\del L\over\del t_{0,Q}}&=&
\biggl\{ \bigl( L^{1/2} \bigr)_+, L\biggr\}~,\CR
 {3\over 2\sqrt2}{\del L \over\del t_{1,Q}}&=&
\biggl\{ \bigl( L^{3/2} \bigr)_+, L \biggr\} ~.
\enn
{}From an examination of lower flow equations we conclude that
the correct normalization of the Hamiltonians is
\eq
(L^n)_+ \Longrightarrow \biggl({1\over\sqrt2}\biggr)^{2n}
{(2n)!! \over (n-1)!}{\del\over\del t_{n-1,R}}~, \quad
(L^{(2n-1)/2})_+ \Longrightarrow \biggl({1\over\sqrt2}\biggr)^{2n-1}
{(2n-1)!! \over (n-1)!}{\del\over\del t_{n-1,Q}}~.
\en

Now let us turn to the problem of the descendant flows of
the puncture operator.
Motivated by the result of the $CP^1$ model \cite{EYCP},  we will look at
the logarithmic Hamiltonians. To define an expansion of $\mbox{log}~L$ in $p$,
we use the following prescription to avoid an appearence of $\mbox{log}~p$;
\eqn
\mbox{log}~L~&=&~{1\over3}~\mbox{log}~
\bigl( 1+ 2 v p^{-1} + 2 wp^{-2} + 2 e^up^{-3}
\bigr) \CR
& &~+{2\over 3}~\mbox{log}~{1\over\sqrt2}\bigl( e^u
+ wp + vp^2 + {1\over 2}p^3 \bigr)~.
\enn
For example,
\eq
\biggl( 3\mbox{log} L\biggr)_-=~
2 vp^{-1} + 2 (w-v^2) p^{-2} + 2\bigl(
e^u + {4\over 3} v^3 - 2 vw
\bigr)p^{-3} + \cdots~.
\en
Contrary to $L^{1/2}$, $\mbox{log} L$ has an infinite series expansion
in both directions of the positive and the negative powers. It is not
appropriate to take a product with $L^{(2n-1)/2}$ or $\mbox{log}~L$,
which is also an infinite series. Hence the next possible generator is
\eqn
\biggl( 3L\mbox{log}L \biggr)_-~&=&~
(2u - \mbox{log}2)e^u p^{-1}
+ \big( e^u + 2 vw -{2\over 3}v^3
\bigr)p^{-1} \CR
& &~~+\biggl( w^2 -2v^2w +{2\over 3}v^4 + 2ve^u
\biggr)p^{-2} \CR
& &~~+ 2
\biggl( (w-v^2)e^u -vw^2
+ {4\over 3}wv^3 - {2\over 5}v^5 \biggr)p^{-3}
 + \cdots~,
\enn
We have found that the flow equations (\ref{pflow0}) and
(\ref{pflow1}) for the descendants of the puncture
operator are recovered from
\eqn
 {2\over 3}{\del L\over\del t_{1,P}}&=&
\biggl\{ \bigl( L( \mbox{log}~L - 1 + {\mbox{log} 2\over 3}) \bigr)_+,
L\biggr\} =
-\biggl\{ \bigl( L( \mbox{log}~L - 1 + {\mbox{log} 2\over 3}) \bigr)_-,
L\biggr\}~, \CR
 {2\over 3}{\del L \over\del t_{2,P}}&=&
\biggl\{  \bigl( L^2( \mbox{log}~L - {3\over 2} + {\mbox{log} 2\over
3})\bigr)_+, L \biggr\} =
-\biggl\{  \bigl( L^2( \mbox{log}~L - {3\over 2} + {\mbox{log} 2\over
3})\bigr)_-, L \biggr\}.
\enn
Thus we have arrived at the following identification;
\eq
\biggl( L^n (\mbox{log} L - c_n)\biggr)_+ \Longrightarrow
{2\over 3}{\del \over \del t_{n,P}}~,
\en
where the constants $c_n$ are determined recursively
by the relation $c_n - c_{n-1} = 1/n$ with the initial
condition $c_0 = -(\mbox{log} 2)/3$.  As a consequence of
this recursion relation, we have a scaling relation;
\eq
{\del\over \del L}L^n(\mbox{log} L - c_n)~=~nL^{n-1}
(\mbox{log} L - c_{n-1})~, \label{logscale}
\en
which is crucial in deriving the string equation.
Finally we observe the relations
\eqn
\langle \sigma_n (P) P \rangle&=& {3\over 2} \biggl(L^n
 ( \mbox{log}~L - c_n)\biggr)_0~, \CR
\langle \sigma_{n-1} (Q) P \rangle&=&(\sqrt 2)^{2n-1}
{(n-1)! \over (2n-1)!!} \biggl(L^{(2n-1)/2} \biggr)_0~, \CR
\langle \sigma_{n-1} (R) P \rangle&=& {1\over n}
\biggl(L^n \biggr)_0~,   \label{res}
\enn
where $(\cdot)_0$ means the degree zero part. These relations
support the validity of the Landau-Ginzburg formalism with $\mbox{log}~p$
as the Landau-Ginzburg variable.

\section{Constraints and String Equation}

\renewcommand{\theequation}{4.\arabic{equation}}\setcounter{equation}{0}

The Lax formalism in section 3 is a special reduction of the Toda
lattice hierarchy. We can reproduce the flow equations
from the Toda lattice hierarchy by imposing constraints on the
Lax operators and the associated Orlov operators.
We should remark, however, that the flows corresponding to
the descendants of the puncture operator are absent in the
standard formulation of the Toda lattice hierarchy.
We will show that the constraints imply the string equation.
Though the type of the constraints is rather different from
the $c=1$ string theory, the method to derive the string
equation is quite similar to it.

Let us first  recall basic ingredients of the (dispersionless)
Toda lattice hierarchy \cite{TTa}.  We consider two Lax operators
$\lax, \overline\lax$ with the following expansion;
\eqn
\lax&=&p + \sum_{n=0}^\infty u_{n+1}(t,\bar t, x)p^{-n}~, \CR
\overline\lax^{-1}&=& \overline u_0 (t,\bar t, x)p^{-1}
+ \sum_{n=0}^\infty \overline u_{n+1} (t,\bar t, x) p^n~.
\enn
The commuting flows are defined by
\eqn
{\del\lax\over\del t_n}&=&\biggl\{ H_n, \lax\biggr\}, \quad
{\del\lax\over\del \overline{t_n}}~=~\biggl\{ \overline H_n, \lax\biggr\},  \CR
{\del\overline\lax\over\del t_n}&=&\biggl\{ H_n, \overline\lax\biggr\},  \quad
{\del\overline\lax\over\del \overline{t_n}}~=~
\biggl\{ \overline H_n, \overline\lax\biggr\},~~(n= 1,2, \cdots)
\enn
where the Hamiltonians are
\eq
H_n~=~(\lax^n)_+~, \qquad \overline H_n~=~(\overline\lax^{-n})_-~.
\en
$X=(X)_+ + (X)_-$ is a decomposition into the non-negative and the negative
power parts. The Poisson bracket $\{\cdot,\cdot\}$ has been introduced
in section 3.
It is convenient to consider the Orlov operators
\eqn
\orl&=&\sum_{n=1}^\infty nt_n\lax^n +x+ \sum_{n=1}^\infty
v_n(t,\bar t,x) \lax^{-n}~, \CR
\overline\orl&=&-\sum_{n=1}^\infty n\overline{t_n}\overline\lax^{-n} +x+
\sum_{n=1}^\infty \overline v_n (t,\bar t,x) \overline\lax^n~. \label{orlov}
\enn
These operators satisfy
\eq
\biggl\{ \lax, \orl \biggr\}~=~\lax, \quad
\biggl\{ \overline\lax, \overline\orl \biggr\}~=~\overline\lax. \label{orldefa}
\en
In fact the above relations together with the following flow equations are
the defining relations of the Orlov opeartors;
\eqn
{\del\orl\over\del t_n}&=&\biggl\{ H_n, \orl\biggr\}, \quad
{\del\orl\over\del \overline{t_n}}~=~\biggl\{ \overline H_n, \orl\biggr\},  \CR
{\del\overline\orl\over\del t_n}&=&\biggl\{ H_n, \overline\orl\biggr\},  \quad
{\del\overline\orl\over\del \overline{t_n}}~=~
\biggl\{ \overline H_n, \overline\orl\biggr\},~~(n=1,2,\cdots) .
\label{orldefb}
\enn

Now it is clear that the Lax operator $L$ introduced in section 3 is obtained
by
imposing the following constraint,
\eq
L~\equiv~{1\over 2}\lax^2~=~\overline\lax^{-1}~. \label{const2}
\en
As a consequence of this constraint we can eliminate the time variables
$\overline{t_n}$ by the identification;
\eq
\overline{t_n}~=~-2^n t_{2n}~.
\en
According to the method of the Riemann-Hilbert problem \cite{TTb}, we should
also
impose the constraint for the conjugate operators;
\eq
-{\cal P}_L~\equiv~\orl\lax^{-2}~=~-\overline\orl~\overline\lax~.
\en
It is the second constraint which leads to the string equation. But before
proceeding to see it, we have to consider a modification of $\orl$ due
to additional flows. As we have seen in section 3, to incorporate the
descendants of the puncture operator, it is necessary to introduce
the flows generated by
\eq
K_n~=~\biggl( L^n (\mbox{log}~L - c_n) \biggr)_+~.
\en
The Hamiltonians $K_n$ can be properly defined only after
imposing the constraint (\ref{const2}), since our prescription to define
$\mbox{log}~L$ requires that the series expansion of $L$
is finite in both the positive and the negative directions.
Let $s_n$ be the time variables for $K_n$.
The Orlov operator is deformed by these new flows, because
it should satisfy;
\eq
{\del\orl\over\del s_n}~=~\biggl\{ K_n, \orl\biggr\}~.
\en
The expression of the modified Orlov operator is
easily obtained by recalling how the defining relations (\ref{orldefa})
and (\ref{orldefb}) imply the expansion (\ref{orlov}).
The original Orlov operator can be reexpressed as
\eq
\orl~=~W (x + \sum_n n t_n p^n) W^{-1}~,
\en
if one introduces a similarity transformation defined by
\eq
\lax~=~WpW^{-1}~.
\en
(We have used a poor notation for the similarity transformation.
It should be understood as the canonical transformation
generated by $\phi$ with $W=e^\phi$. For a mathematically
more precise definition, see \cite{TTb}.)
Using the flow equation for  $W$;
\eq
{\del W\over \del t_n}~=~H_n W -W p^n~,
\en
one can check the defining relations are indeed satisfied.
(We neglect $\bar t_n$ flows, since they have been eliminated
by the constraint.) Defining the coefficient functions $\{ v_n\}$
by
\eq
WxW^{-1}~=~x + \sum_{n=1}^\infty v_n \lax^{-n}~,
\en
we get the expansion (\ref{orlov}).
Now the $s_n$ flows for $W$ are given by
\eq
{\del W\over \del s_n}~=~K_n W -W \biggl[ \big({p^2\over 2}\big)^n
\big(\mbox{log}{p^2\over 2} - c_n\big)\biggr]~.
\en
We see that
\eq
\orl'~=~W \biggl(x + \sum_n n t_n p^n + \sum_n n s_n {p^{2n}\over 2^{n-1}}
(\mbox{log} {p^2\over 2} - c_{n-1})\biggr) W^{-1}~,
\en
satisfies the additional flow equations. Here we have used
the scaling property (\ref{logscale}).
Hence we obtain the following expansion of the modified Orlov
operator
\eq
\orl'~=~x + \sum_n n t_n \lax^n + \sum_n 2n s_n L^n
(\mbox{log}~L - c_{n-1} ) + \sum_n v_n \lax^{-n}~.
\en

Let  us return to a derivation of the string equation.
The second constraint allows us to express
the conjugate operator in a linear combination of the
Hamiltonians. To see it, it is convenient to consider the non-negative
power part and the negative power part separately.
Since $\lax^{-n}~(n\geq 1)$ does not have any positive power part and
$\overline\lax^n~(n\geq 1)$ does not have any negative power part,
we have
\eqn
({\cal P}_L)_+ &=&(-\orl' \lax^{-2})_+ = -\sum_{n \geq 3} nt_n H_{n-2}
- \sum_{n\geq 1} ns_n K_{n-1}~,\CR
({\cal P}_L)_-&=& (\overline\orl~\overline\lax)_-
=- \sum_{n\geq 2} n \overline{t_n} \overline H_{n-1}~.
\enn
Hence ${\cal P}_L$ is expressed as
\eq
-{\cal P}_L~=~2\sum_{n\geq 2} 2n t_{2n} H_{2n-2} + \sum_{n\geq 1} (2n+1)
t_{2n+1}
H_{2n-1} + \sum_{n\geq 1} n s_n K_{n-1}~. \label{combi}
\en
We have used the first constraint to eliminate $\overline{t_n}$
and $\overline H_{n-1}$.
Combined with the canonical commutation relation $\{ {\cal P}_L, L \}=1$,
(\ref{combi}) implies,
\eq
2\sum_{n\geq 2} 2n t_{2n} {\del L \over \del t_{2n-2}}
+ \sum_{n\geq 1} (2n+1) t_{2n+1} {\del L \over \del t_{2n-1}}
+ \sum_{n\geq 1} n s_n {\del L \over \del s_{n-1}}~=~-1~.
\en
After an appropriate identification of the flow pamameters as the coupling
parameters and the shift $s_1 \longrightarrow s_1 -1$ of the dilaton coupling,
we obtain the string equation, or the puncture equation;
\eq
1+ \sum_{n\geq 1} nt_{n,R}{\del L \over \del t_{n-1,R}}
+ \sum_{n\geq 1} nt_{n,Q}{\del L \over \del t_{n-1,Q}}
+ \sum_{n\geq 1} nt_{n,P}{\del L \over \del t_{n-1,P}}~=~
{\del L \over \del t_{0,P}}~.
\en

It is easy to generalize the above argument to the model with
the constraint
\eq
L~\equiv~{\lax^N \over N}~=~\overline \lax^{-1}~, \quad (N = 3, 4, \cdots).
\en
The Lax operator takes the form
\eq
L~=~{1\over N}p^N + v_{N-1}p^{N-1} + \cdots + v_0 + e^u p^{-1}~.
\en
Since the Lax operator has a finite expansion in both directions,
we can define an expansion of $\mbox{log}~L$ in a similar way to the case
$N=2$. The string equation will follow from the associated constraint;
\eq
{\cal P}_L~=~\orl \lax^{-N}~=~-\overline\orl~\overline\lax~.
\en

\section{Higher Genus Expansion}

\renewcommand{\theequation}{5.\arabic{equation}}\setcounter{equation}{0}

\newcommand {\R}{R_{n}^{'}}
\newcommand {\RR}{R_{n}^{''}}
\newcommand {\RRR}{R_{n}^{'''}}
\newcommand {\w}{w_{n}^{'}}
\newcommand {\ww}{w_{n}^{''}}
\newcommand {\www}{w_{n}^{'''}}
\newcommand {\wwww}{w_{n}^{''''}}
\renewcommand {\u}{u_{n}^{'}}
\newcommand {\uu}{u_{n}^{''}}
\newcommand {\uuu}{u_{n}^{'''}}
\newcommand {\uuuu}{u_{n}^{''''}}
\renewcommand {\v}{v_{n}^{'}}
\newcommand {\vv}{v_{n}^{''}}
\newcommand {\vvv}{v_{n}^{'''}}
\newcommand {\vvvv}{v_{n}^{''''}}

To obtain the genus expansion of the flow eqations, we promote
the dispersionless Lax operator into the $N\times N$ matrix
Lax operator;
\eqn
Q&=&\sum_n E_{n,n+1} + \sum_n\sum_{l\geq 0} a_n (\ell) E_{n,n-\ell}~,
\label{Q1} \\
Q^2&=&\sum_n E_{n,n+2} + \sum_n \sqrt 2 \widetilde v_n E_{n,n+1} +
 \sum_n w_n E_{n,n} +  \sum_n {1\over\sqrt 2} R_n E_{n,n-1}~, \label{Q2}
\enn
where $E_{i,j}$ stands for the matrix unit.
The degree of $E_{i,j}$ is defined to be $(j-i)$. That is, the positive degree
part
is the upper triangular part of the matrix. The reason we add the tilde
on $v_n$ will become clear shortly.
Though we will not introduce an explicit matrix model realization,
we believe the matrix Lax operator $Q$ would arise from
an appropriate multi-matrix model developed, for example,
in \cite{BX2}.

The commuting flows are defined by
\eqn
{n\over N}{\del Q \over \del t_{n-1,R}}&=& \biggl[ (Q^{2n})_+, Q \biggr]~,\\
{1\over N} \biggl({1\over\sqrt 2}\biggr)^{2n-1}{(2n-1)!! \over (n-1)!}
{\del Q \over \del t_{n-1,Q}}&=& \biggl[ (Q^{2n-1})_+, Q \biggr]~,\\
{2\over 3N}{\del Q \over \del t_{n,P}}&=&
\biggl[ \bigl(Q^{2n} (\mbox{log}~Q^2 - c_n)\bigr)_+, Q \biggr]~,
\enn
where $[\cdot,\cdot ]$ is the commutator of the matrices.
The form of the flow equations is conformable to the genus
expansion only when we make a good choice of (independent) dynamical
variables. Since it is the free energy of the theory that should
be expanded in genus by genus, two point functions are
\lq\lq good" variables.  From the residue formula (\ref{res}) in section 3
the diagonal matrix elements are identified with
two point functions in full genus expansion.
Thus we take $w_n$ as one of the variables, but not $\widetilde v_n$.
Instead we can choose $a_n(0)$ as the second variable.
Therefore, in the following we will use a notation $v_n \equiv a_n(0)$.
$\widetilde v_n$ coincides with $v_n$ in the dispersionless limit.
As the last variable, motivated by the matrix  model method,
we take $\phi_n$ with $R_n = \exp N(\phi_n - \phi_{n-1})$.
The functions $\mbox{log}~\phi_n$ correspond to the weight functions for
the orthogonal polynomials in the matrix model.

The genus expansion can be obtained in the following way.
We first write down the flow equations for the matrix elements.
Then we make a Taylor expansion of $w_{n+k}, v_{n+\ell}$
and $R_{n+m}$ around $w_n, v_n$ and $R_n$, respectively,
with an identification $x\equiv n/N$. The continuum limit is
defined to be the limit where the matrix size $N$ goes to the infinity.
We thus obtain the flow equations in terms of differential polynomials
in $w_n, v_n$ and $\phi_n$. (As before the prime
denotes the derivative with respect to $x$.) The order of $N^{-1}$
practically counts the number of the derivatives.
Regarding $N^{-1}$ as an expansion parameter or the
cosmological constant, we can see the higher
genus structure of the flow equations.

Let us illustrate the above procedure for the $t_{n,R}$-flows.
The primary flow equations in terms of the matrix elements are
\eqn
\frac{\partial {{\phi}}_n}{\partial t_{0,R}} &=&
{{w}}_n~, \\
\frac{1}{N}
\frac{\partial {{v}}_{n}}{\partial t_{0,R}} &=&
{{R}}_{n+1}-{{R}}_n~, \\
\frac{1}{N} \frac{\partial {{w}}_n}{\partial t_{0,R}} &=&
\frac{1}{2} \bigl[ ({{v}}_n +{{v}}_{n+1})
 {{R}}_{n+1} -({{v}}_{n-1}+{{v}}_n )
{{R}}_n \bigr] .
\enn
Similarly the $t_{1,R}$-flow equations are given by
\eqn
\frac{\partial \phi_n}{\partial t_{1,R}} &=&
\frac{1}{2} w_n^2 +\frac{1}{4}\bigl[ (v_n +v_{n+1})R_{n+1} +(v_{n-1} +v_n
)R_{n} \bigr]~, \\
 \frac{1}{N}\frac{\partial v_n}{\partial t_{1,R}} &=&\frac{1}{2} (w_{n}+
w_{n+1})R_{n+1}
 -\frac{1}{2}(w_{n-1}+w_n)R_n~, \\
\frac{1}{N} \frac{\partial w_n}{\partial t_{1,R}} &=&
 \frac{1}{4} (w_n +w_{n+1})(v_n +v_{n+1})
R_{n+1} -\frac{1}{4}(w_{n-1}+w_n)(v_{n-1} +v_n )R_n \CR & &
+ \frac{1}{4}(R_{n+1}R_{n+2} -R_{n-1}R_n ) .
\enn
Making a Taylor expansion of $w_{n \pm 1}, v_{n\pm 1}, R_{n+1}$ and $R_{n+2}$,
we get the genus expanded flow equations. Up to genus two the primary flows
read
\eqn
\frac{\partial u_n}{\partial t_{0,R}} &=& w_n ' ~,\\
\frac{\partial v_n}{\partial t_{0,R}} &=&
\biggl[ e^{u_n}+ \frac{1}{24N^2}(u_{n}^{'2} +2\uu )e^{u_{n}} \CR
 & &+ \frac{1}{N^4}(\frac{\uuuu}{360} +\frac{\u \uuu}{180} + \frac{7u_{n}^{'2}
\uu}{1440}
+\frac{u_{n}^{'4}}{1920} + \frac{u_{n}^{''2}}{240})e^{u_n}\biggr]' ~,\\
\frac{\partial w_n}{\partial t_{0,R}} &=& \biggl[ ve^{u_n} +
\frac{1}{24N^2}(4\vv +2\u \v +
u_{n}^{'2} v_n +2\uu v_n)e^{u_n} \CR
& &+ \frac{1}{N^4}\biggl(\frac{\vvvv}{120}
+\frac{v_n \uuuu}{360} +\frac{\u \vvv}{80} + \frac{\v \uuu +v_n \u \uuu}{180}
+ \frac{11\uu \vv}{720} \CR
& &+ \frac{u_{n}^{'2} \vv}{120} +\frac{v_n u_{n}^{''2}}{240} + \frac{7 v_n
u_{n}^{'2} \uu}{1440} +
\frac{u_{n}^{'3} \v }{480} +\frac{v_n u_{n}^{'4}}{1920}
+ \frac{7\v \u \uu}{720}\biggr)e^{u_n}\biggr]'~,
\enn
where we have introduced $u_n = \phi'_n$.
The $t_{1,R}$-flows are quite lengthy and we collect them in Appendix B.
We think it is quite non-trivial
that the odd-power terms in $N^{-1}$ disappear from the flow equations
in accord with our identification of $N^{-1}$ as the genus expansion parameter.
This fact supports our prescription of genus expansion and choice of
dymanical variables.

Computation of $t_{n,Q}$-flows is more complicated, since the flow equations
involve
the matrix elements $a_n(\ell)$. Identifying the square of (\ref{Q1}) with
(\ref{Q2}), we have to eliminate these variables. We present only the final
results up to genus two;
\eqn
\frac{\partial u_n}{\partial t_{0,Q}} &=& v_n ' \\
\frac{\partial v_n}{\partial t_{0,Q}} &=& \biggl[ w_n -\frac{v_{n}^{2}}{2} -
\frac{1}{12N^2}(\ww -v_n \vv -v_{n}^{'2} ) \CR
& &+ \frac{1}{120N^4}(\wwww -v_n \vvvv +4\v \vvv -7v_{n}^{''2} )\biggr]' \\
\frac{\partial w_n}{\partial t_{0,Q}} &=&
\biggl[ e^{u_n} + \frac{1}{24N^2} ({u_n '}^2 +2u_n ^{''})e^{u_n} \CR
& &+ \frac{1}{N^4}(\frac{\uuuu}{360} +\frac{\u \uuu}{180} + \frac{7u_{n}^{'2}
\uu}{1440}
+\frac{u_{n}^{'4}}{1920} + \frac{u_{n}^{''2}}{240})e^{u_n}\biggr]'
\enn
Note that $\del w_n / \del t_{0,Q} = \del v_n / \del t_{0,R}$ as it should be.
The $t_{1,Q}$-flows are again collected in Appendix B.

Finally let us look at the $t_{1,P}$-flow as an example of flow
equations involving $\mbox{log}~Q^2$.
In terms of the matrix element of $\mbox{log}~Q^2$, the flow equations are
given by;
\eqn
\frac{2}{3} \frac{\partial \phi_n}{\partial t_{1,P}} &=&
\frac{R_n}{\sqrt{2}}b_{n-1} +w_n a_n +\sqrt{2} \widetilde v_n d_{n+1} +e_{n+2}
\\
\frac{2}{3N} \frac{\partial \widetilde v_n}{\partial t_{1,P}} &=&
\widetilde v_n \biggl( \frac{R_n b_{n-1} -R_{n+1} b_n}{\sqrt{2}}
+w_n a_n -w_{n+1} a_{n+1} \CR & &+ \sqrt{2}(\widetilde v_n d_{n+1}
-\widetilde v_{n+1} d_{n+2})+e_{n+2} -e_{n+3} \biggr)\CR
& &+ (w_{n+1} -w_n ) \left(\frac{R_n}{2}c_{n-1} + \frac{w_n}{\sqrt{2}}b_n
+\widetilde v_n a_{n+1} +\frac{d_{n+2}}{\sqrt{2}} \right) \CR
 & &+ \frac{R_{n+2}}{2} \left(\frac{R_n }{\sqrt{2}}f_{n-1} +w_n c_n +
\sqrt{2} \widetilde v_n b_{n+1} + a_{n+2} \right) \CR
& &- \frac{R_n}{2} \left(\frac{R_{n-1}}{\sqrt{2}}f_{n-2} + w_{n-1} c_{n-1}
+\sqrt{2} \widetilde v_{n-1} b_{n} +a_{n+1}\right) \\
\frac{2}{3N} \frac{\partial w_n}{\partial t_{1,P}} &=&
 R_{n+1} \biggl( \frac{R_n}{2}c_{n-1} +\frac{w_n}{\sqrt{2}}b_n +\widetilde v_n
a_{n+1} +
\frac{d_{n+2}}{\sqrt{2}} \biggl) \CR
& &- R_n \biggl( \frac{R_{n-1}}{2}c_{n-2} + \frac{w_{n-1}}{\sqrt{2}}b_{n-1}
+\widetilde v_{n-1} a_n +\frac{d_{n+1}}{\sqrt{2}} \biggr)
\enn
where we have used the following notations for the matrix elements
of $\mbox{log}~Q^2$;
\eqn
a_n &:=& (\mbox{log}~Q^2)_{n,n} ,\ \ \ \ b_n :=(\mbox{log}~Q^2)_{n,n+1} ,
\ \ \ \ c_n := (\mbox{log}~Q^2)_{n,n+2} \CR
f_n &:=& (\mbox{log}~Q^2)_{n,n+3}, \ \ \ \ d_n := (\mbox{log}~Q^2)_{n,n-1} ,
\ \ \ \ e_n :=(\mbox{log}~Q^2)_{n,n-2} .
\enn
and omitted the counter terms proportional to $(3-\mbox{log} 2)/3$.  We have
used
$\widetilde v_n$ for simplicity of the expressions.
The computation of the matrix elements of $\mbox{log}~Q^2$ is rather
technical and given in Appendix C. Substituting it and making a Taylor
expansion,
we get the $t_{1,P}$-flow equations as follows;
\eqn
\frac{\partial u_n }{\partial t_{1,P}} &=&
\biggl[ u_n w_n + \frac{v_{n}^{2}}{2}
+\frac{1}{24N^2}(6\ww -3v_{n}^{'2} -2v_n \vv )\biggr]' ~,\\
\frac{\partial v_n }{\partial t_{1,P}} &=&
\biggl[ (u_n -1)e^{u_n} +v_n w_n -\frac{v_{n}^{3}}{3}
+\frac{1}{24N^2}((5u_{n}^{'2} + 6\uu +u_n u_{n}^{'2} +2u_n \uu )e^{u_n } \CR
& & +6v_n v_{n}^{'2} +4 v_{n}^2 \vv -4v_n \ww -2\v \w )\biggr]' ~,\\
\frac{\partial w_n }{\partial t_{1,P}} &=& \biggl[ \frac{w_{n}^{2}}{2} + u_n
v_n e^{u_{n}}
+ \frac{e^{u_n }}{24N^2}(4v_n u_{n}^{'2} +v_n u_n u_{n}^{'2}+ 6\u \v +2u_n \u
\v \CR
& & +6v_n \uu +2v_n u_n \uu +6\vv +4u_n \vv ) \biggr]'~,
\enn
where we have calculated up to genus one.
As we show in Appendix A, the formula of one loop free energy
by Dijkgraaf and Witten implies the flow equations at one loop
level. In this section we have computed several lower flow equations
up to genus two (genus one for $t_{1,P}$) by
making expansion of flow equations in matrix form.
One loop order of our result enjoys a complete agreement
with what is derived from the formula of one loop free energy.

The existence and consistency of higher genus expansion
seems to impose very severe restrictions on possible models
of topological string theory \cite{EYY}.
For example, the $A_2$ minimal model and the $CP^1$ model exhaust
the two primary model with a consistent genus expansion.
There are other topological matter theories, for instance, based on the Lie
algebra $B_2 =C_2$. However, this model does not have a consistent higher
genus expansion beyond genus one.
It is believed that the generalized KdV hierarchy controls the genus expansion
of the minimal topological string theory of ADE type in the sense that
the genus expansion of the free energy comes from a single tau
function of the integrable hierarchy. In this paper we have found
another series of topological strings with
a well-defined genus expansion, starting from the topological
$CP^1$ model. The Toda lattice hierarchy plays
the same role as the KP hierarchy for the minimal models.
In this respect, to obtain more complete understanding of higher genus
behavior,
it is highly desirable to find a matrix model realization of the model, which
will
provide a closed expression of the genus expansion of the free energy.

\bigskip

We would like to thank T.Eguchi, K.Takasaki,
Y.Yamada and S.-K.Yang for discussions.
The work of H.K. is supported in part
by the Grant-in-Aid for Scientific Reseach from
the Ministry of Eduaction, Science and Culture
(No.06221255 and No.06740069).

\begin{center}
\bf Appendix A : One Loop Free Energy
\end{center}

\renewcommand{\theequation}{A.\arabic{equation}}\setcounter{equation}{0}

There is a formula of Dijkgraaf-Witten for the one-loop free energy
of topological string \cite{Wit}\cite{DW}. We can derive the flow euations at
one loop level
from the one-loop free enegy together with the tree level
flow equations\footnote{We owe the material in this section
to Y.~Yamada.}. To see it in general model, let ${\cal O}_\alpha,
{\cal O}_\beta, \cdots$ be the primary fields and $t_{n,\alpha}$
the flow parameter for $\sigma_n ({\cal O}_\alpha)$.
The flow equation for $u_\alpha = \langle P{\cal O}_\alpha\rangle$ in full
genus
is assumed to be
\eq
{\del u_\alpha\over \del t_{n,\beta}}~=~\biggl[ R_{\alpha,\beta,n}^{(0)}
+ \lambda^2 R_{\alpha,\beta,n}^{(1)} + \cdots \biggr]'~,
\en
where $R_{\alpha,\beta,n}^{(g)} = R_{\alpha,\beta,n}^{(g)} [u_\gamma]$ is a
potential
at genus $g$.
It is important to remember that the order parameter $u_\gamma$ has to be
expanded as
\eq
u_\gamma~=~u_\gamma^{(0)} + \lambda^2 u_\gamma^{(1)} + \cdots~.
\en
What we need is the potential $R_{\alpha,\beta,n}^{(1)}[u_\gamma^{(0)}]$.
The sub-leading part of the full genus flow equation gives
\eq
{\del u_\alpha^{(1)} \over \del t_{\beta,n}}~=~\biggl[ \sum_\gamma
{\delta R_{\alpha,\beta,n}^{(0)}\over \delta u_\gamma} u_\gamma^{(1)} +
R_{\alpha,\beta,n}^{(1)}[u_\gamma^{(0)}] \biggr]'~.
\en
Now, in terms of the one-loop free energy;
\eq
F_1~=~{1\over 24} \mbox{log det}~\langle P{\cal O}_\alpha {\cal O}_\beta
\rangle_0~,
\en
we have
\eq
u_\gamma^{(1)}~=~{\del^2 F_1 \over \del t_{0,P} \del t_{0,\gamma}}~.
\en
Hence we obtain the potential
\eq
R_{\alpha,\beta,n}^{(1)}[u^{(0)}]~=~{\del^2 F_1 \over \del t_{0,\alpha} \del
t_{n,\beta}}
- \sum_\gamma {\delta R_{\alpha,\beta,n}^{(0)} \over \delta u_\gamma}
{\del^2 F_1 \over \del t_{0,P} \del t_{0,\gamma}}~.
\en
The existence of the second term is crutial for obtaining the potential in
a form of differential polynomial in $u_\gamma^{(0)}$. The first term gives a
rational
expression in general. We have computed one loop potentials for our model
with three primaries using the above formula and found
a complete agreement with the flow equations in section 5 (See also Appendix
B).

\begin{center}
\bf Appendix B :  Flow Equations up to Genus Two
\end{center}

\renewcommand{\theequation}{B.\arabic{equation}}\setcounter{equation}{0}
\begin{center}
\underline{ $t_{1,R}$-flows }
\end{center}
\eqn
\frac{\partial u_n}{\partial t_{1,R}} &=&
\biggl[ v_n e^{u_n} +\frac{{w_n}^2} {2} + \frac{e^{u_n}}{24N^2}(6v_{n}^{''}
+6u_n ' v_n ' +4u_{n}^{''} v_n +3u_{n}^{'2} v_n ) \CR
& &+ \frac{e^{u_n}}{5760N^4} \biggl(15v_n u_{n}^{'4} +120v_n u_{n}^{'2}
\uu +120v_n \u \uuu +120 \uuu \v +120 \vvvv \CR
& &+ 80v_n u_{n}^{''2} +48v_n \uuuu +60u_{n}^{'3} \v +240\u \uu \v
+240\uu \vv + 240\u \vvv \CR & &+ 180u_{n}^{'2} \vv \biggr)\biggr]' \\
\frac{\partial v_n}{\partial t_{1,R}} &=&
\biggl[ w_n e^{u_n} + \frac{e^{u_n}}{24N^2}(4\ww +2\u \w + u_{n}^{'2} w_n
+2\uu w_n) \CR
& &+ \frac{e^{u_n}}{5760N^4}\biggl(48\wwww +16w_n \uuuu +72\u w_{n}^{'''}
 + 32\w u_{n}^{'''} +32w_n \u u_{n}^{'''} \CR
& &+ 88\uu \ww + 48u_{n}^{'2} \ww +24w_{n} u_{n}^{''2}
+ 28 w_n u_{n}^{'2} \uu +12 u^{'3} \w +3w_n u_{n}^{'4} \CR
& &+ 56\w \u \uu\biggr)\biggr]' \\
\frac{\partial w_n}{\partial t_{1,R}} &=&
\biggl[ w_n v_n e^{u_n} + \frac{1}{2} e^{2u_n} \CR
& &+ \frac{1}{24N^2} \bigl((2v_n w_n \uu +4v_n \ww +4w_n \vv
+ w_n v_n u_{n}^{'2} +2 \v \w \CR & &+ 2v_n \w \u +2w_n \v \u )e^{u_n} +
8(\uu +u_{n}^{'2})e^{2u_n}\bigr) \CR
& &+ \frac{1}{5760N^4} \biggl( (384u_{n}^{'4} +768u_{n}^{''2}
 + 256\uuuu +1024 \u \uuu +1792 u_{n}^{'2} \uu)e^{2u_n} \CR
 & &+ (96\u \w \vv +96\u \v \ww
+72w_n \u \vvv +72\w \vvv +72v_n \u \www +72\v \www \CR
 & &+ 48w_n u_{n}^{'2} \vv +48w_n \vvvv +48v_n \wwww
+48v_n u_{n}^{'2} \ww +32v_n w_n \u \uuu \CR
 & &+ 24v_n w_n u_{n}^{''2} +36\v \w u_{n}^{'2} + 16v_n w_n \uuuu
+ 12w_n \v u_{n}^{'3} +12v_n \w u_{n}^{'3} \CR
& &+ 56 w_n \u \v \uu +56v_n \u \w \uu
+56 \v \w \uu + 88w_n \uu \vv + 88v_n \uu \ww \CR
 & &+28v_n w_n u_{n}^{'2} \uu
+ 3v_n w_n u_{n}^{'4} )e^{u_n} \biggr) \biggr]'
\enn
\newpage
\begin{center}
\underline {$t_{1,Q}$-flows }
\end{center}
\eqn
\frac{\partial u_n }{\partial t_{1,Q}} &=& \biggl[ e^{u_n} +v_n w_n
- \frac{v_{n}^{3}}{6} \CR & &+ \frac{1}{24N^2}\bigl((5 u_{n}^{'2} +6\uu
)e^{u_n}
+4v_n v_{n}^{'2} -2\v \w +2v_{n}^{2} \vv -2v_n \ww \bigr) \CR
& &+ \frac{1}{5760N^4}\biggl( (5u_{n}^{'4} +20u_{n}^{'2} \uu
+40u_{n}^{''2}+80\u \uuu +48\uuuu) e^{u_n} \CR
& &- 360\vv v_{n}^{'2} -360v_n v_{n}^{''2} + 120\v \www
 -600v_n \v \vvv +120v_n \wwww \CR
& &-120v_{n}^{2} \vvvv\biggr) \biggr]' \\
\frac{\partial v_n }{\partial t_{1,Q}} &=&
\biggl[ v_n e^{u_{n}} + \frac{w_{n}^{2}}{2} -\frac{v_{n}^{2} w_n}{2}
+\frac{v_{n}^{4}}{8} + \frac{1}{24N^2}\bigl((6\vv +4\u \v -v_n
u_{n}^{'2})e^{u_n } \CR
 & &- 6v_{n}^{2} v_{n}^{'2} +2w_n v_{n}^{'2} + 2w_n v_n \vv -2w_n \ww
+4v_n \v \w +3v_{n}^{2} \ww -3v_{n}^{3} \vv - w_{n}^{'2}\bigr) \CR
& &+\frac{1}{5760N^4}\biggl( (63v_n u_{n}^{'4}+ 72\v u_{n}^{'3}
+368v_n \uu u_{n}^{'2} +112\v \uuu +256\v \u \uu \CR & &
+184v_n u_{n}^{''2} +88\uu \vv +232v_n \u \uuu +56v_n \uuuu
+ 28 u_{n}^{'2} \vv+32\u \vvv+48\vvvv)e^{u_n} \CR & & +264v_{n}^{'4}
+2184v_n \vv v_{n}^{'2} -456\v \vv \w +1128v_{n}^2 \v \vvv
+816v_{n}^2 v_{n}^{''2} \CR
& &-48w_n v_n \vvvv +48w_n \wwww -
144w_n v_{n}^{''2}-144v_{n}^2 \wwww +144v_{n}^3 \vvvv -368 v_{n}^{'2} \ww \CR
& & +112 \w \www -488v_n \vv \ww + 104w_{n}^{''2} -472v_n \v \www
 - 192w_n \v \vvv \CR
& & -192v_n \vvv \w \biggr)\biggr]' \\
\frac{\partial w_n}{\partial t_{1,Q}} &=&
 \biggl[ (w_n +\frac{v_{n}^{2}}{2} )e^{u_n} \CR
 & &+ \frac{e^{u_n}}{24N^2}(2v_n \vv -2v_{n}^{'2} +2\u \w + 6\ww
+2v_n \v \u +\frac{v_{n}^{2} u_{n}^{'2}}{2} +w_n u_{n}^{'2} \CR
& &+ v_{n}^{2} \uu +2w_n \uu )
+ \frac{e^{u_n}}{5760N^4} (64v_{n}^{''2} -64\u \v \vv
+ 48v_n \vvvv +48\wwww \CR
& &+48v_n \vv \uu
+ 32\v \vvv +32v_n \vvv \u -32v_{n}^{'2} \uu +32v_n \v \uuu +32\w \uuu \CR
& &+32w_n \u \uuu + 112\www \u +16v_{n}^{2} \u \uuu +16w_n \uuuu
+ 12v_n \v u_{n}^{'3} -12v_{n}^{'2} u_{n}^{'2} \CR & &+12\w u_{n}^{'3} +12
v_{n}^2 u_{n}^{''2}+ 28v_n \vv u_{n}^{'2} +68\ww u_{n}^{'2}
+ 28w_n u_{n}^{'2} \uu +\frac{3v_{n}^{2} u_{n}^{'4}}{2} \CR
 & &+3w_n u_{n}^{'4} +128\uu \ww
+ 56\w \u \uu +56v_n \v \u \uu +8v_{n}^{2} \uuuu \CR
 & &+ 14v_{n}^{2} \uu u_{n}^{'2} +24w_n u_{n}^{''2})\biggr]'
\enn

\newpage
\begin{center}
\bf Appendix C : Matrix Elements of $\mbox{log}~Q^2$
\end{center}
\renewcommand{\theequation}{C.\arabic{equation}}\setcounter{equation}{0}

\newcommand {\A}{Y_0}
\newcommand {\B}{\sqrt 2 z\widetilde v_{n}^{(1)} +w_{n}^{(1)}
+\frac{1}{\sqrt{2}} z^{-1} R_{n}^{(1)} }
\newcommand {\C}{\sqrt 2 z\widetilde v_{n}^{(2)} +w_{n}^{(2)}
+\frac{1}{\sqrt{2}} z^{-1} R_{n}^{(2)} }
\newcommand {\D}{\sqrt 2 z\widetilde v_{n}^{(m)} +w_{n}^{(m)}
+\frac{1}{\sqrt{2}} z^{-1} R_{n}^{(m)} }
\newcommand {\E}{Y_{1}^{(1)}}
\newcommand {\F}{Y_{1}^{(2)}}
\newcommand {\G}{Y_{1}^{(3)}}
\newcommand {\z}{z\partial}
\renewcommand {\v}{\widetilde v_{n}^{'}}
\renewcommand {\vv}{\widetilde v_{n}^{''}}
\renewcommand {\vvv}{\widetilde v_{n}^{'''}}
\renewcommand {\vvvv}{\widetilde v_{n}^{''''}}

To evaluate the matrix elements of $\mbox{log}~Q^2$, we use the following
formula
\footnote{We are grateful
to S.-K.~Yang for telling us a computation in the $CP^1$ model.}.
The $(n,m)$ element is given by
\eqn
(\mbox{log} Q^2)_{n,m} &=& \frac{1}{2\pi i} \oint \frac{dz}{z} z^{n-m}
\mbox{log} (z^2 +\sqrt{2}z\widetilde v(n+\z) +w(n+\z) +\frac{1}{\sqrt{2}}
z^{-1}R(n+\z)) \cdot 1~, \CR
&\equiv& \frac{1}{2\pi i} \oint \frac{dz}{z} z^{n-m}
\mbox{log}~Y(z) .
\enn
Taylor expansion gives
\eq
Y(z) ~=~{\A}(z) +\sum_{m=1}^{\infty} \frac{1}{m!} ({\D})\cdot (\z)^m \cdot 1~,
\en
where
\eq
\A (z) ~=~ z^2 +\sqrt{2}z\widetilde v_n +w_n
+\frac{1}{\sqrt 2}z^{-1}R_n~,
\en
and $f^{(m)}$ denotes the $m$-th derivative of $f$.
Then $\mbox{log}~Y(z)$ may be expanded as
\eq
\mbox{log}~Y(z) ~=~ \sum_{m=1}^{\infty} \frac{(-1)^{m-1}}{m} [(\A
-1)+\sum_{n=1} ^{\infty}
\frac{1}{n!} Y_{1}^{(n)} (\z)^{n} ]^m ~,
\en
where
\eq
Y_{1}^{(m)} ~=~ \D .
\en
Let us organize the expansion by the number of the derivatives;
\eq
\mbox{log}~Y(z)~=~\mbox{log}~Y_0(z) + A_1 + A_2 + A_3 + \cdots,
\en
where $A_n$ denotes the terms including $n$ derivatives.
We need the contributions up to $A_3$ to obtain the one loop flow
equations. The first order term, for example, is computed as follows.
$[(\A -1) + \E \z]^m \cdot 1$ gives terms like
\eq
(\A -1)^{m-n} \E \z (\A -1)^{n-1} ~=~ (n-1)(\A -1)^{m-n}(\A -1)^{n-2} \E
\dot{\A}~,
\en
where the dot means $z(d / dz)$.  Using the formulae
\eqn
\sum_{m=1}^{n} m(m+1) \cdots (m+k) &=& \frac{1}{k+2} \frac{(n+k+1)!}{(n-1)!}
{}~,\\
\sum_{n=1}^{\infty} (-1)^{n-1}
(\A -1)^{n-1} &=& \frac{1}{\A}~,
\enn
we obtain
\eqn
A_1 &=& \sum_{m=1}^{\infty} \frac{(-1)^{m-1}}{m} \sum_{n=1}^{m-1}
 (\A -1)^{m-n} \E \z (\A -1)^{n-1} \CR
&=& \frac{1}{2} \sum_{m=1}^{\infty}
(-1)^{m-1} \frac{d}{d\A} (\A -1)^{m-1} \E \dot{\A} \CR
&=& -\frac{1}{2{\A}^2} \E \dot{\A}.
\enn
Similar computation gives
\eq
A_2 ~=~-\frac{1}{4{\A}^2} \F \ddot{\A} + \frac{1}{3{\A}^3}
 \bigl(\F \dot{\A^{2}} +\E {\dot{Y_{1}}}^{(1)} \dot{\A} +(\E)^2 \ddot{\A}\bigr)
- \frac{3}{4{\A}^4}(\E)^2 \dot{\A^{2}}
\en
\eqn
A_3 &=& -\frac{1}{12} \frac{\G}{Y_{0}^{2}} {\stackrel{...}{Y}}_{0}
+\frac{1}{3} \frac{\G}{Y_{0}^{3}} {\dot{Y}}_{0} {\ddot{Y}}_{0}
+ \frac{1}{6} \frac{\F}{Y_{0}^{3}} \left(z\frac{d}{dz} \right)^2 \left(\E
{\dot{Y}}_{0}\right)
+ \frac{1}{6} \frac{\E}{Y_{0}^{3}} z \frac{d}{dz} \left( \F
{\ddot{Y}}_{0}\right) \CR
& & -\frac{1}{4} \frac{\G}{Y_{0}^{4}} {\dot{Y}}_{0}^{3}
-\frac{3}{8} \frac{\E \F}{Y_{0}^{4}} {\dot{Y}}_{0} {\ddot{Y}}_{0}
- \frac{1}{4} \frac{\E}{Y_{0}^{4}} z \frac{d}{dz} \left(\F {\dot{Y}}_{0}^2
\right) \CR
& &-\frac{3}{4} \frac{\F {\dot{Y}}_{0}}{{Y}_{0}^{4}} z \frac{d}{dz} \left(\E
{\dot{Y}}_{0} \right)
- \frac{1}{4} \frac{\E}{{Y}_{0}^{4}} z\frac{d}{dz}
\left( \E z \frac{d}{dz} \left(\E {\dot{Y}}_{0} \right) \right) \CR
& & +2 \frac{\E \F}{Y_{0}^{5}} {\dot{Y}}_{0}^{3}
+ 2 \frac{(\E )^2 {\dot{Y}}_{0}}{{Y}_{0}^{5}} z\frac{d}{dz} \left( \E
{\dot{Y}}_{0}\right)
-\frac{5}{2} \frac{{\E}^3 ({\dot{Y}}_{0})^3} {{Y}_{0}^{6}} .
\enn

Now, we can evaluate the matrix elements $(\mbox{log}~Q^2)_{n,m}$.
The residue integral picks up the coefficient of $z^{m-n}$
in a Laurent expansion of $\mbox{log}~Y(z)$.
We have to be careful about the point where the expansion is made.
In accord with our prescription of defining $\mbox{log}~L$ for the
dispersionless
Lax operator, we add two contributions, one from the origin and the other
from the infinity in an appropriate weight;
\eq
(\mbox{log}~Q^2)_{n,m}~=~\frac{2}{3}c_{n,m}^{(0)} +
\frac{1}{3}c_{n,m}^{(\infty)} ~.
\en
With this prescription we obtain
\eqn
(\mbox{log}~Q^2)_{n,n} &=& \frac{2}{3} \mbox{log} \frac{R_{n}}{\sqrt{2}} +
\frac{\R}{3R_n N} + \frac{\RR}{18R_n N^2} -\frac{R_{n}^{'2}} {18R_{n}^{2} N^2}
-\frac{\R \RR}{4R_{n}^{2} N^3} \\
\frac{1}{\sqrt{2}}(\mbox{log} Q^2)_{n,n+1} &=&
\frac{2w_n }{3R_n } +\frac{\w R_n -2w_n \R} {3R_{n}^{2} N} \CR
& &+ \frac{1}{18R_{n}^{3}N^2}
(\ww R_{n}^2 -6w_n R_n \RR +12w_n R_{n}^{'2} -6 \w \R R_n ) \CR
& &+ \frac{1}{36R_{n}^{4}N^3}\biggl[ -24w_n R_{n}^{'3} + 60w_n R_n \R
\RR -4R_{n}^2 w_n \RRR \CR
& & + 12R_n R_{n}^{'2} \w -15 R_{n}^2 \RR \w - 11R_{n}^2 \R \ww \biggr] \\
\frac{1}{2}(\mbox{log} Q^2)_{n,n+2} &=& \frac{2\widetilde v_n}{3R_n} -
\frac{w_{n}^{2}}{3R_{n}^{2}} -
\frac{1}{3R_{n}^{3}N}(2w_n \w R_n +3\widetilde v_n R_n \R -R_{n}^2 \v -
3w_{n}^{2} \R ) \CR
& &+ \frac{1}{18R_{n}^{4}N^2} (15w_{n}^{2} R_n \RR +30\widetilde v_n R_n
R_{n}^{'2}
+36w_n \w \R R_n \CR
& &- 6w_n \ww R_{n}^{2} -17\widetilde v_n R_{n}^2 \RR -10\v \R R_{n}^{2}
-5w_{n}^{'2} R_{n}^2
+ \vv R_{n}^{3} -42w_{n}^{2} R_{n}^{'2}) \CR
& &+ \frac{1}{36R_{n}^{5}N^3} \biggl[180w_{n}^{2} R_{n}^{'3}
+ 36\v R_{n}^{'2} R_{n}^2 +30R_{n}^2 \R w_{n}^{'2} + 156\widetilde v_n R_{n}^2
\R \RR \CR
& &+ 96w_n \w \RR R_{n}^2 +72w_n \ww \R R_{n}^2 +18w_{n}^{2} R_{n}^2 \RRR
- 29\v \RR R_{n}^3 \CR
& &-13\vv \R R_{n}^3 -17\w \ww R_{n}^3 - 24\widetilde v_n R_{n}^3 \RRR
 -4w_n \www R_{n}^3
-108\widetilde v_n R_n R_{n}^{'3} \CR & &- 168w_n \w R_{n}^{'2} R_n -234w_{n}^2
R_n
 \R \RR \biggr]\\
\frac{1}{\sqrt{2}}(\mbox{log} Q^2)_{n,n+3} &=&
\frac{2}{3R_n} -\frac{4\widetilde v_n w_n}{3R_{n}^{2}}+
\frac{4w_{n}^{3}}{9R_{n}^{3}} \CR
& &+ \frac{2}{3R_{n}^{4}N}(-2R_{n}^2 \R -2\v w_n R_{n}^2
-3\w \widetilde v_n R_{n}^2 \CR
& &- 4w_{n}^{3} \R +3w_{n}^{2} \w R_n +8\widetilde v_n w_n R_n \R ) \CR
 & &+ \frac{1}{9R_{n}^5 N^2}
\biggl[ -14\v \w R_{n}^3 -17R_{n}^3 \RR - 6\vv w_n R_{n}^3 -17\ww \widetilde
v_n R_{n}^3
+ 28 R_{n}^2 R_{n}^{'2} \CR
& &+48w_n \v \R R_{n}^2+76\widetilde v_n \w \R R_{n}^2 +
24w_n w_{n}^{'2} R_{n}^2 +60\widetilde v_n w_n R_{n}^2 \RR +15w_{n}^{2} \ww
R_{n}^2 \CR
& &- 28w_{n}^{3} R_n \RR -108w_{n}^{2} \w \R R_n - 156\widetilde v_n w_n R_n
R_{n}^{'2}
+ 100w_{n}^{3}R_{n}^{'2} \biggr] \CR
 & &+ \frac{1}{18R_{n}^6 N^3} \biggl[ -720w_{n}^{3} R_{n}^{'3}
+ 960\widetilde v_n w_n R_n R_{n}^{'3} +900w_{n}^{2} \w R_{n}^{'2} R_n \CR
 & &+660w_{n}^{3} R_n \R \RR - 144R_{n}^2 R_{n}^{'3}
+112\widetilde v_n w_n R_{n}^3 \RRR+18w_{n}^{2} \www R_{n}^3 \CR
& & -312w_n \v R_{n}^{'2} R_{n}^2 -516\widetilde v_n \w R_{n}^{'2} R_{n}^2
- 288w_n w_{n}^{'2} \R R_{n}^2 \CR
& &- 996\widetilde v_n w_n R_{n}^2 \R \RR -342w_{n}^{2} \w \RR R_{n}^2
-270w_{n}^{2}
\ww \R R_{n}^2 -48w_{n}^{3} R_{n}^2 \RRR \CR
& &+ 120\v \w \R R_{n}^3 +18w_{n}^{'3} R_{n}^3 + 177R_{n}^3 \R \RR
+156w_n \v \RR R_{n}^3 \CR & &+ 232\widetilde v_n \w \RR R_{n}^3
+ 84w_n \vv \R R_{n}^3 + 184\widetilde v_n \ww \R R_{n}^3 + 108w_n \w \ww
R_{n}^3 \CR
& &- 19\w \vv R_{n}^4 +31\v \ww R_{n}^4 - 36R_{n}^4 \RRR -4 \vvv w_n R_{n}^4
-24
\www \widetilde v_n R_{n}^4 \biggr]\\
\frac{1}{\sqrt{2}}(\mbox{log} Q^2)_{n,n-1} &=&
\frac{\widetilde v_n}{3} - \frac{\v}{3N} +\frac{\vv}{9N^2} \\
 (\mbox{log}~Q)_{n,n-2} &=& \frac{w_n }{3}
-\frac{\widetilde v_{n}^{2}}{3}
- \frac{\w}{3N} +\frac{\widetilde v_n \v}{N} +\frac{4\ww -11\widetilde v_n
\vv -12\widetilde v_{n}^{'2}}{18N^2} \CR
& & +\frac{8}{3N^3} \v \vv +\frac{\widetilde v_n \vvv}{6N^3} .
\enn

\newcommand{\npb}[1]{{\it Nucl.~Phys.}~{\bf B#1}}
\newcommand{\mpla}[1]{{\it Mod.~Phys.~Lett.}~{\bf A#1}}
\newcommand{\cmp}[1]{{\it Commun.~Math.~Phys.}~{\bf #1}}
\newcommand{\plb}[1]{{\it Phys.~Lett.}~{\bf B#1}}
\newcommand{\ptp}[1]{{\it Prog.~Theor.~Phys.}~{\bf #1}}
\newcommand{\ijmp}[1]{{\it Int.~J.~Mod.~Phys.}~{\bf A#1}}
\newcommand{\lmp}[1]{{\it Lett.~Math.~Phys.}~{\bf #1}}

\end{document}